\documentclass[twocolumn,superscriptaddress,letter]{revtex4}
\usepackage{amssymb}
\usepackage{color}
\usepackage{graphicx}
\usepackage{dcolumn}
\usepackage{bm}
\usepackage{amsmath}
\usepackage{amsfonts}
\usepackage[colorlinks,
            linkcolor=blue,
            anchorcolor=blue,
            citecolor=blue
            ]{hyperref}
\begin{document}

\title{Defective Majorana zero modes in non-Hermitian Kitaev chain}
\author{Xiao-Ming Zhao}
\affiliation{Beijing National Laboratory for Condensed Matter Physics, Institute of
Physics, Chinese Academy of Sciences, Beijing 100190, China}
\author{Cui-Xian Guo}
\affiliation{Beijing National Laboratory for Condensed Matter Physics, Institute of
Physics, Chinese Academy of Sciences, Beijing 100190, China}
\affiliation{Center for Advanced Quantum Studies, Department of Physics, Beijing Normal
University, Beijing 100875, China}
\author{Su-Peng Kou}
\affiliation{Center for Advanced Quantum Studies, Department of Physics, Beijing Normal
University, Beijing 100875, China}
\author{Lin Zhuang}
\affiliation{ State Key Laboratory of Optoelectronic Materials and Technologies, School of Physics, Sun Yat-Sen University, Guangzhou 510275, China}
\author{Wu-Ming Liu}
\thanks{Corresponding author}
\email{wliu@iphy.ac.cn }
\affiliation{Beijing National Laboratory for Condensed Matter Physics, Institute of
Physics, Chinese Academy of Sciences, Beijing 100190, China}
\affiliation{Songshan Lake Materials Laboratory, Dongguan, Guangdong 523808, China}

\begin{abstract}
Topological stability is an important property for topological materials. However, the non-Hermitian effects may change this situation. Here, we investigate the robustness of edge states in the non-Hermitian Kitaev chain with imbalanced tunneling term and superconducting
pairing term. By defining the similarity of Majorana zero modes (MZMs) and magnetic factor, the coalescing phase diagram of the MZMs and corresponding spin polarization phase diagram are provided. Because of the non-Hermitian coalescence effect and non-Hermitian suppression effect induced by the breakdown of sublattice symmetry and particle-hole symmetry, the system emergence very interesting phenomenons, such as defective MZMs, number-anomalous bulk-boundary correspondence, coalescing of many-body ground states, the magnetic phase crossover without gap closing. Those novel non-Hermitian effects offer fresh insights into MZMs and topological physics.
\end{abstract}

\maketitle

\section{Introduction}
As a prototype model of one-dimensional (1D)
topological superconductors (SCs), Kitaev chain have been a hot
spot in condensed matter physics since unpaired Majorana zero modes (MZMs) are
predicted to exist at the ends of this chain when the system
is in the topologically nontrivial phase \cite{kitaev2001}, which is robust for perturbation. Due to the
potential applications in topological quantum computation (TQC),
Majorana fermions or MZMs have been widely studied in
recent years, including finding MZMs in different materials \cite{Fu2008,Mourik2012,Deng2012,Rokhinson2012,Alicea2012,Mebrahtu2013,Nadj-Perge2014,Lee2014}, its non-Abelian statistics \cite{read2000,Ivanov2001,Sarma2006,Stern2010} and its application in TQC \cite{Tewari2007,Nayak2008,Sau2010,Alicea2011}.

Recently, non-Hermitian (NH) physics attracts lots of attention \cite{Ashida2020}. A non-Hermitian Hamiltonian has been introduced to describe the NH open system, which is regarded as a subsystem of an infinite Hermitian system \cite{Subsystem}. The NH systems present many novel topological properties, such as exceptional points (EPs) in complex energy spectrum \cite{Yin2018,Ghatak2019,EPReview2021}, anomalous topological transition \cite{Rudner2009,Esaki2011,Hu2011,Shen2018,Lieu2018,Gong2018,Yin2018, Jiang2018,Ghatak2019,38-1,38,chen-class2019, Kunst2019} and edge states \cite{Lee2016,Leykam2017,KawabataUeda2018,kou2020}. One of the most important properties for the topological systems is the bulk-boundary correspondence (BBC), i.e., bulk topological invariants of the bulk energy spectrum can predict unique gapless boundary states. Due to the NH skin effect \cite{Yao2018,YaoWang2018,SongWang2019,Deng2019,Longhi2019}, the typical BBC is broken, it obeys the non-Block BBC relationship \cite{Xiong2018,Kunst2018,Jin2019,Lee2019,Herviou2019,Yokomizo2019}. Experimentally, those topological properties of the NH systems have been investigated in platform of photonics \cite{Weimann2017, Zeuner2015,Bandres2018,Zhou20182, Cerjan2019}, quantum walks \cite{Xiao2017,Wang2019,Xiaoxue2019} and electric circuits \cite{Helbig2019}.

While, it is still an open question whether the topological robustness of MZMs can be changed by the NH effects in superconducting systems. Physically, an NH Kitaev chain can be obtained by setting the the chemical potential or hopping amplitude or SC paring to complex values, these
choices may be realized by adding onsite particle gain/loss \cite{Wang2015,San2016,Yuce2016,Zeng2016,Menke2017,Kawabata2018,Lieu2019}, or introducing nonreciprocal effects to the nearest neighbor hopping amplitude \cite{YaoWang2018,Yao2018}, or imbalanced tunneling of particle paring \cite{Li2018,SongZhi,ImblanceParing,Unpair2019}, respectively.
 In previous research about the MZMs in NH Kitaev chain,  the BBC is no different from the Hermitian cases both for the located (gain/loss) and dislocated(imbalanced SC paring) NH effects. Experimentally, the Hermitian Kitaev model can be realized in nanowire \cite{HemitianKC} and the related PT-symmetric many-body system have been presented in Ref.\cite{PTKC}, the nonreciprocal effects can also be realized by using a cavity array with passive nearest neighbor tunneling \cite{Array}. Specifically, the NH Kitaev model with imbalanced particle paring may
be realized by loading fermionic cold atoms in a 1D optical
lattice, where the effective p-wave pairing can be induced by
an optical Raman transition \cite{PwaveParing}, and the non-Hermiticity
may be implemented by controlling and monitoring the decay
of atoms \cite{NHKC1,NHKC2,NHKC3,NHKC4}.

In this letter, we find that the breakdown of chiral symmetry and particle-hole (PH) symmetry can induce defective MZMs, which means one of the two localized edge states will disappear, referred to as number-anomalous of the MZMs. As a result, the conventional BBC is broken down, this indicates the NH effects break the robustness of the topological Majorana bound states. Here, the NH effects make difference are the coalescence effects and NH suppression effect. The coalescence effect means the similarity rate of two states $\gamma=\left\langle\psi_{1}|\psi_{2}\right\rangle$ changes from 0 to 1, i.e., two orthogonal states coalesce into one in a nonunitary way; For a target state $\psi=\alpha_{1}\left|\uparrow\right\rangle+\alpha_{2}\left|\downarrow\right\rangle$, the NH suppression effect means the NH parameters change the population of $\left|\uparrow\right\rangle$ and $\left|\downarrow\right\rangle$ in a non-unitary way.

This paper is organized as follows. In Sec.\ref{sectII}, we describe the model Hamiltonian and analyze its topological invariant. In Sec.\ref{DefMajorana}, we investigate the MZMs based on effective edge states Hamiltonian in the Majorana representation, and the number-anomalous BBC is revealed. In Sec.\ref{ManyBody}, we study the NH effects on the defective MZMs in spin language and show the phase crossover in the corresponding NH Ising model. Finally, we provide a summary and discussion in Sec.\ref{Summary}.

\begin{figure}[t]
\centering
\includegraphics[clip,width=0.5\textwidth]{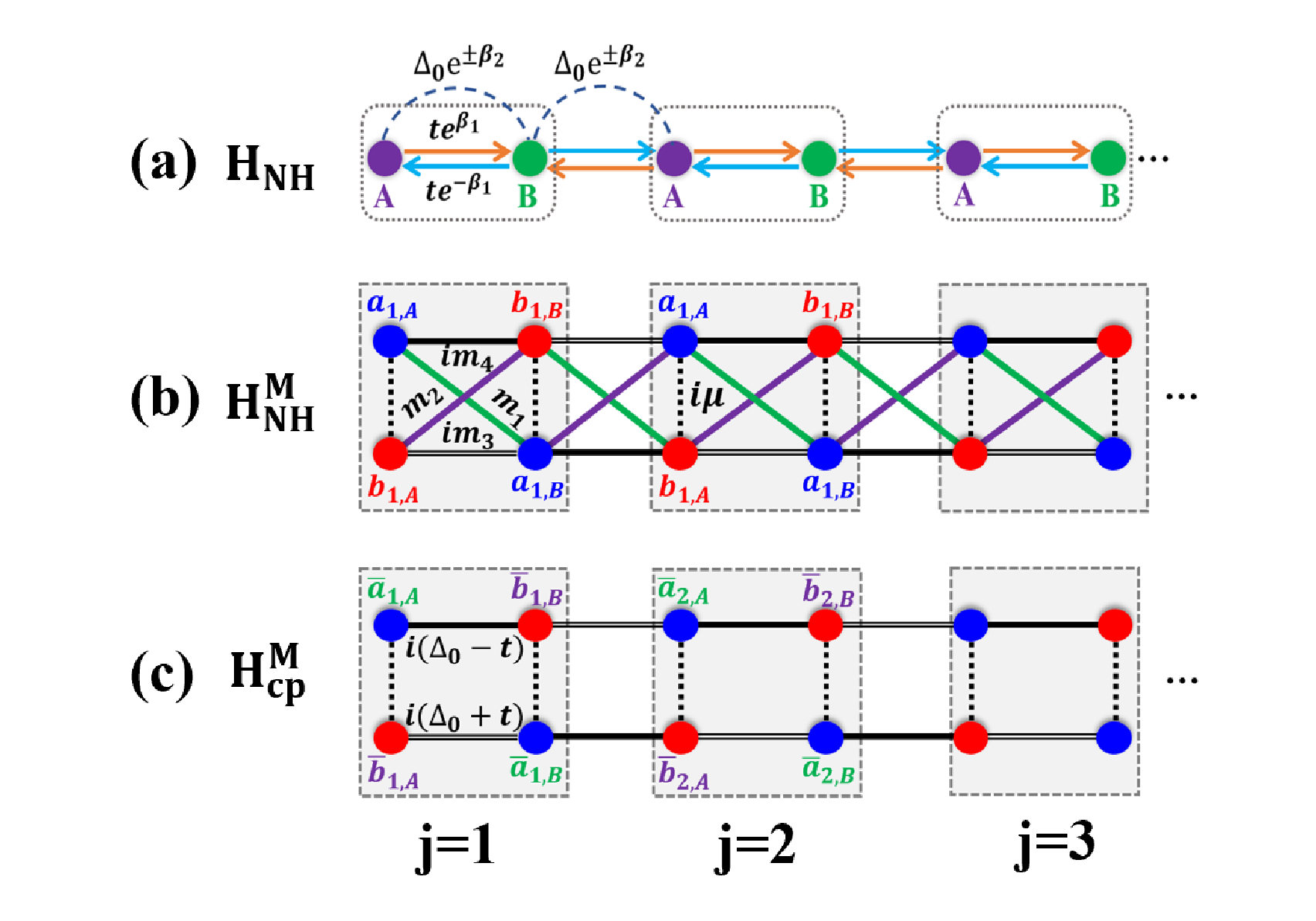}
\caption{ Schematic illustration of the non-Hermitian Kitaev chain in different representation, the dotted box indicates the unit cell. \textbf{(\textbf{a})} The lattice structure in Dirac Fermion representation, which is described by $\mathrm{H}_{\mathrm{NH}}$. The imbalanced hopping amplitudes from A to B (orange arrow) and from B to A (blue arrow) are described by $\beta_{1}$; The imbalanced paring amplitudes of particles and holes are described by $\beta_{2}$. \textbf{(\textbf{b})} Kitaev chain viewed as two coupled SSH chains in Majorana representation, which is described by $\mathrm{H}_{\mathrm{NH}}^{\mathrm{M}}$. The Majorana Fermion $a_{j,A/B}$ $(b_{j,A/B})$ are marked by blue (red) filled circle. The solid lines indicate the couplings $m_{1},m_{2},im_{3},im_{4}$ between nearest neighbor sites, and the dashed lines indicate the couplings $i\mu$ intra-site. \textbf{(\textbf{c})} The lattice schematic diagram of the Hermitian counterpart systerms in Majorana representation, which is described by $\mathrm{H}_{\mathrm{cp}}^{\mathrm{M}}$. }
 \label{chain}
\end{figure}

\section{Non-Hermitian Kitaev chain and topological invariant}\label{sectII}

\subsection{Hamiltonian of non-Hermitian Kitaev model}

We introduce a 1D NH Kitaev model induced by the breakdown of chiral symmetry and particle-hole symmetry, the Hamiltonian is
\begin{eqnarray}
H_{\mathrm{NH}} &=&-\sum_{j}[t_{L}c_{j}^{\dagger }c_{j+1}+t_{R}c_{j+1}^{\dagger
}c_{j}  \notag \\
&&+\Delta^{+}c_{j}^{\dagger }c_{j+1}^{\dagger }+\Delta^{-}c_{j+1}c_{j}+\mu (1-2n_{j})],
\label{Hamilt}
\end{eqnarray}
where $c_{j}^{\dagger}(c_{j})$ is a fermionic creation (annihilation) operator on site $j$. As shown in Fig.\ref{chain}\textcolor[rgb]{0.00,0.00,1.00}{(a)}, we introduce the imbalanced hopping strength $\beta_{1}$ and the imbalanced SC paring strength $\beta_{2}$ in the system, which has two sublattice in a unit cell. $t_{L/R}=te^{\pm\epsilon\beta_{1}}$ denote the left/right hopping amplitude where $\epsilon=\pm1$ for inter/intra cell. $\Delta^{\pm}=\Delta_{0}e^{\pm\beta_{2}}$ is the amplitude of $p$-wave pair creation (annihilation), $\mu$ is the chemical potential.

Then, we rewrite $H_{\mathrm{NH}}$ in the Bogoliubov-de Gennes formalism $H_{\mathrm{NH}}=C^{\dagger }h_{\mathrm{BdG}}C$ where $C$ and $C^{\dagger}$ are column and row vectors containing all canonical operators
\begin{eqnarray}
C&=&(c_{1,A},\cdots,c_{N,B},c_{1,A}^{\dagger },\cdots,c_{N,B}^{\dagger})^{T},\nonumber\\
C^{\dagger}&=&(c_{1,A}^{\dagger },\cdots,c_{N,B}^{\dagger},c_{1,A},\cdots,c_{N,B}).
\label{NambRep}
\end{eqnarray}
then we can transform it to its Hermitian counterpart using a similarity transformation
\begin{eqnarray}
h_{\mathrm{cp}}=S_{2}S_{1}h_{\mathrm{BdG}}S_{1}^{-1}S_{2}^{-1}
\end{eqnarray}
where the transform matrices are defined as
\begin{eqnarray}
S_{1}&=&diag\{1,r_{1},\cdots,1,r_{1},1,r_{1}^{-1},\cdots ,1,r_{1}^{-1}\},\nonumber \\
S_{2}&=&diag\{r_{2},\cdots ,r_{2},r_{2}^{-1},...,r_{2}^{-1}\}
\end{eqnarray}
and $r_{1}=\mathrm{exp}(-\beta _{1})$, $r_{2}=\mathrm{exp}[(\beta _{1}-\beta _{2})/2]$.
So the Hamiltonian of the Hermitian counterpart read as \cite{SongZhi,Li2018}
\begin{eqnarray}
H_{\mathrm{cp}}=-\sum_{j}\{(t d_{j+1}^{\dagger }d_{j}+\Delta _{0}d_{j}^{\dagger }d_{j+1}^{\dagger }+h.c.)+\mu (1-2n_{j})\},
\label{Hcp}
\end{eqnarray}
where the canonical operators are defined as
\begin{eqnarray}
d_{j}\equiv \Omega _{j}c_{j},d_{j}^{\dagger }\equiv \Omega_{j}^{-1}c_{j}^{\dagger }
\label{simTran}
\end{eqnarray}
and the scale factors of similar transformation are
\begin{eqnarray}
\Omega _{j\in odd}=e^{(\beta _{1}-\beta _{2})/2},\Omega _{j\in even}=e^{-(\beta _{1}+\beta _{2})/2}.
\end{eqnarray}
It is obvious that the operators satisfy the  anti-commutation relations $\{d_{i},d_{j}^{\dagger }\}=\delta _{ij}$, $\{d_{i},d_{j}\}=\{d_{j}^{\dagger
},d_{i}^{\dagger }\}=0$ and $H_{\mathrm{cp}}$ has the form of the 1D Hermitian Kitaev model.

By the Fourier transform, We can write down the Bogoliubov-de Gennes Hamiltonian in the momentum space as
\begin{equation}
H_{\mathrm{NH}}(k)=-\frac{1}{2}C^{\dag }h(k)C
\end{equation}%
where $C$ and $C^{\dag }$ are column and row vectors containing all canonical operators, $C=(c_{k,A},c_{k,B},c_{-k,A}^{\dag
}$, $c_{-k,B}^{\dag })^{T},C^{\dag }=(c_{k,A}^{\dag },c_{k,B}^{\dag
},c_{-k,A},c_{-k,B})$, and the non-Hermitian matrix is
\begin{equation}
h(k)=\left(
\begin{array}{cccc}
-2\mu & e^{-\beta _{1}}K_{+} & 0 & e^{\beta _{2}}X_{+} \\
e^{\beta _{1}}K_{-} & -2\mu & -e^{\beta _{2}}X_{-} & 0 \\
0 & -e^{-\beta _{2}}X_{+} & 2\mu & -e^{\beta _{1}}K_{+} \\
e^{-\beta _{2}}X_{-} & 0 & -e^{-\beta _{1}}K_{-} & 2\mu%
\end{array}%
\right)
\end{equation}
where $K_{\pm }=t(1+e^{\pm ik})$, $X_{\pm }=\Delta _{0}(1-e^{\pm ik})$.
Diagonalizing $h(k)$, we can get the energy dispersion
\begin{equation}
E(k)=\pm 2\sqrt{(t\cos(k)-\mu)^{2}+\Delta^{2}\sin^{2}(k)},
\end{equation} which can't be
affected by the non-Hermitian parameter $\beta _{1},\beta _{2}$ and is just
the spectrum of $\beta _{1}=\beta _{2}=0$. It should be noticed that there is no skin effect for the system \cite{Yao2018,YaoWang2018,SongWang2019,Deng2019,Longhi2019}. Therefore, one can calculate the gap closing point $E(k)=0$ to get the topological phase boundary of the system. As a result, the Majorana edge modes emerge in the topological phase $|\mu/t|<1$.

\begin{figure}[t]
\scalebox{0.35}{\includegraphics*[0.5in,0.0in][10.7in,5.8in]{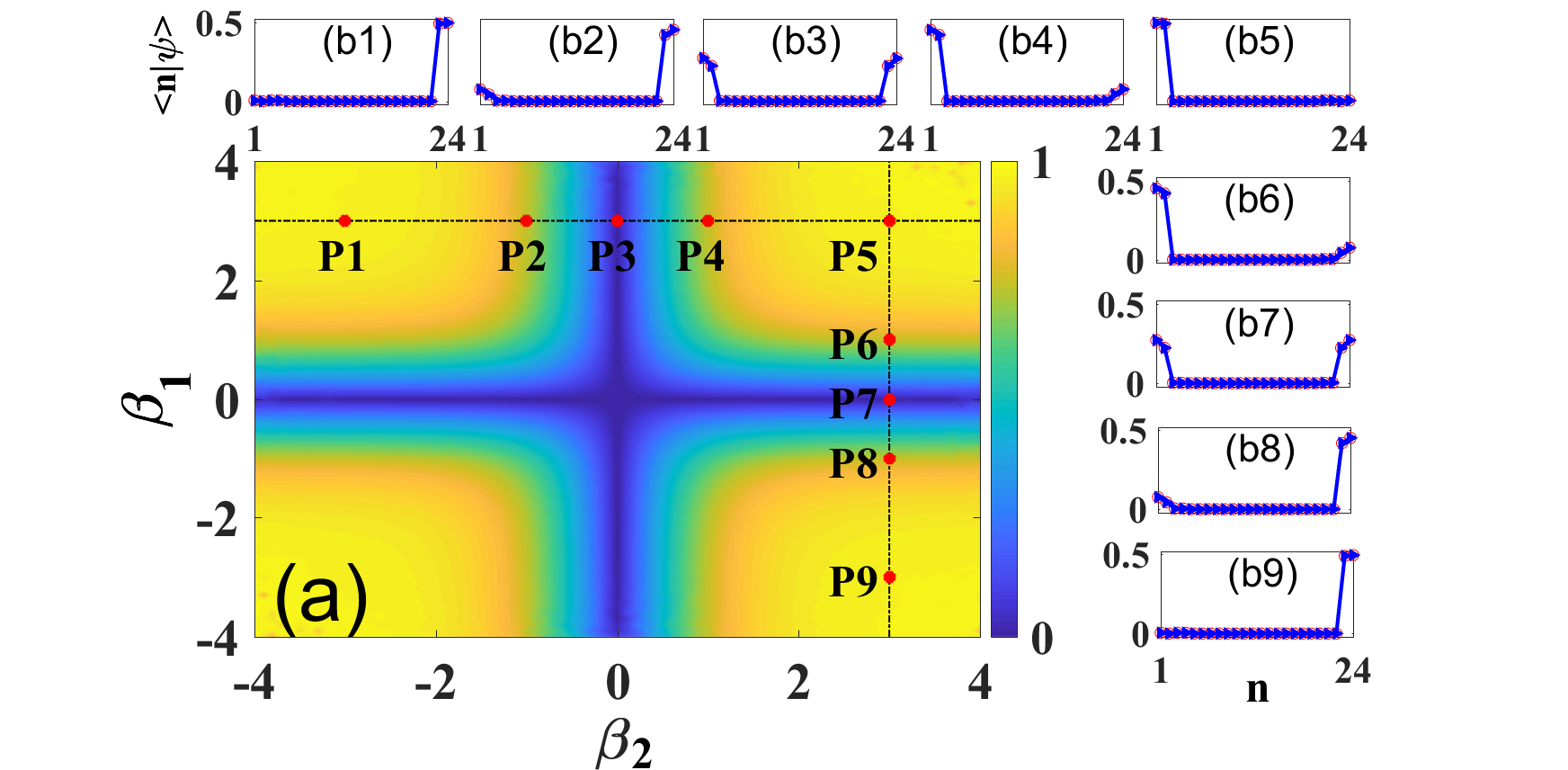}}
\caption{Similarity $\gamma _{\mathrm{M}}$ and localized properties of the two defective MZMs with $L=24$, $\mu=0.1$, $t=1$ and $\Delta_0=0.8$.
\textbf{(a)} The blue area represent $\gamma _{\mathrm{M}}=0$ which means two MZMs local at two end of the chain independently, the yellow areas represent $\gamma _{\mathrm{M}}=1$ which means the two MZMs coalesce to one and local at either left or right end of the chain.
\textbf{(b1-b9)} The distributions of the MZMs for different cases marked by the points P1-P9 in (a), respectively. For the cases in P1, P5, P9, which are away from $\beta_{1}=0$ and $\beta_{2}=0$, the two MZMs coalesce to one.}
 \label{PhaseDiagramMajorana}
\end{figure}

\subsection{Biorthogonal $\mathrm{Z}_{2} $ topological invariant}
Then, we investigate the topological properties of the NH Kitaev chain $H_{\mathrm{NH}}$. The fermion Hamiltonian in momentum space is divided into three parts,
\begin{equation}
\mathrm{\hat{H}}_{\mathrm{NH}}=\mathrm{\hat{H}}_{\mathrm{NH},k>0}%
+\mathrm{\hat{H}}_{\mathrm{NH},k=0}+\mathrm{\hat{H}}_{\mathrm{NH},k=\pi}.
\end{equation}
After diagonalizing the fermion Hamiltonian at the points $k>0$, we have
\begin{equation}
\mathrm{\hat{H}}_{\mathrm{NH},k>0}=\sum_{k\mathbf{>}0}\varepsilon
(k)\alpha_{k}^{\dag}\alpha_{k}-\sum_{k>0}\varepsilon(k)\alpha_{-k}\alpha
_{-k}^{\dagger}%
\end{equation}
where $\alpha_{k}$ are diagonalized quasi-particles operators and $\alpha_{\pm
k}$ annihilate the ground state $|G \rangle$, i.e., $\alpha_{\pm k}|G
\rangle=0$. Both $\alpha$ band has a positive energy at each point in momentum
space $k>0$. The Hamiltonian at $k=0$ and $k=\pi$ are diagonalized into
\begin{align}
\mathrm{\hat{H}}_{\mathrm{NH},k=0}  &  =\varepsilon(k=0)\alpha_{k=0}^{\dag
}\alpha_{k=0},\nonumber \\
\mathrm{\hat{H}}_{\mathrm{NH},k=\pi}  &  =\varepsilon(k=\pi)\alpha_{k=\pi
}^{\dag}\alpha_{k=\pi}%
\end{align}
where $\varepsilon(k=0)=\mu-t$ and $\varepsilon(k=\pi)=\mu+t.$

Based on the biorthogonal set, the right/left eigenstates and corresponding eigenvalues for the NH systems satisfy the relationship $\mathrm{\hat{H}}_{\mathrm{NH}}|{\Psi}_{m}^{\mathrm{R}}%
\rangle=E_{m}|{\Psi}_{m}^{\mathrm{R}}\rangle,$ and $\mathrm{\hat{H}%
}_{\mathrm{NH}}^{\dagger}|{\Psi}_{m}^{\mathrm{L}}\rangle=E_{m}^{\ast}|{\Psi
}_{m}^{\mathrm{L}}\rangle$. To describe this topological structure of $\mathrm{\hat{H}}_{\mathrm{NH}}$,
we define \emph{biorthogonal }$Z_{2}$\emph{ topological invariant},
\begin{equation}
\mathcal{\omega}=\mathrm{sgn}(\eta_{_{k=0}}\cdot \eta_{_{k=\pi}})
\end{equation}
where
\begin{align}
\eta_{_{k=0}}  &  =\left \langle {\Psi}_{0}^{\mathrm{L}}\right \vert
c_{k=0}^{\dagger}c_{k=0}\left \vert {\Psi}_{0}^{\mathrm{R}}\right \rangle
{},\nonumber \\
\eta_{_{k=\pi}}  &  =\left \langle {\Psi}_{0}^{\mathrm{L}}\right \vert c_{k=\pi
}^{\dagger}c_{k=\pi}\left \vert {\Psi}_{0}^{\mathrm{R}}\right \rangle .
\end{align}
Here, $\left \vert {\Psi}_{0}^{\mathrm{R}}\right \rangle $ denotes the ground
state, $\Delta_{0}=t$, and
\begin{align}
\eta_{_{k\mathbf{=}0}}  &  =\frac{\varepsilon \lbrack k=0 \mathbf{]}%
}{\left \vert \varepsilon \lbrack k=0 \mathbf{)]}\right \vert }=\frac{\mu-t
}{\left \vert \mu-t \right \vert },\nonumber \\
\eta_{_{k\mathbf{=}\pi}}  &  =\frac{\varepsilon \lbrack k=\pi \mathbf{]}%
}{\left \vert \varepsilon \lbrack k=\pi \mathbf{)]}\right \vert }=\frac{t+\mu
}{\left \vert t+\mu \right \vert }.
\end{align}
As a results, we have
\begin{align}
\eta_{_{k=0}}  &  =\{%
\begin{array}
[c]{l}%
+1,\text{ }\varepsilon \left(  k=0\right)  >0\\
-1,\text{ }\varepsilon \left(  k=0\right)  <0
\end{array}
,\nonumber \\
\eta_{_{k=\pi}}  &  =\{%
\begin{array}
[c]{l}%
+1,\text{ }\varepsilon \left(  k=\pi \right)  >0\\
-1,\text{ }\varepsilon \left(  k=\pi \right)  <0
\end{array}
.
\end{align}
Then $\mathcal{\omega}$ becomes topological invariant to characterize the
universal properties of different topological phases, $\mathcal{\omega}=1$
with $(\eta_{k=0},\eta_{k=\pi})=\left(  -1,-1\right)  $ or $\left(
1,1\right)  $ represent two trivial SCs, and $\mathcal{\omega}=-1$ with
$(\eta_{k=0},\eta_{k=\pi})=\left(  -1,1\right)  $ or $\left(  1,-1\right)  $
represent two topological SCs, where there exist two MZMs located at two end of the 1D Kitaev chain.

\section{Defective Majorana edge states}\label{DefMajorana}
\subsection{Hamiltonian in Majorana representation}
By introduce Majorana Fermion $a_{n}=c_{n}^{\dag}+c_{n}$, $b_{n}=-i(c_{n}^{\dag}-c_{n})$, the non-Hermitian Hamiltonian  $H_{\mathrm{NH}}$ can be written in Majorana-representation:
\begin{eqnarray}
H^{\mathrm{M}}_{\mathrm{NH}}
=&-&\frac{1}{4}\sum_{j}[m_{1}(a_{j,A}a_{j,B}-b_{j,B}b_{j+1,A})\nonumber \\
&+&m_{2}(b_{j,A}b_{j,B}-a_{j,B}a_{j+1,A})\nonumber \\
&+& im_{3}(b_{j,A}a_{j,B}+b_{j,B}a_{j+1,A})\nonumber \\
&+& im_{4}(-a_{j,A}b_{j,B}-a_{j,B}b_{j+1,A})\nonumber \\
&+& i4\mu (a_{j,A}b_{j,A}+a_{j,B}b_{j,B})],
\end{eqnarray}
where the coupling $m_{1},m_{2},im_{3},im_{4}$ between nearest neighbor sites are:
\begin{eqnarray}
m_{1}&=&t\sinh(\beta_{1})+\Delta_{0}\sinh(\beta_{2}),\nonumber \\
m_{2}&=&t\sinh(\beta_{1})-\Delta_{0}\sinh(\beta_{2}),\nonumber \\
m_{3}&=&t\cosh(\beta_{1})+\Delta_{0}\cosh(\beta_{2}),\nonumber \\
m_{4}&=&-t\cosh(\beta_{1})+\Delta_{0}\cosh(\beta_{2}).
\end{eqnarray}

As shown in Fig.\ref{chain}\textcolor[rgb]{0.00,0.00,1.00}{(b)}, the system contains four Majorana Fermions in each unit cell. the Majorana Fermion $a_{j,A/B}$ $(b_{j,A/B})$ are marked by blue (red) filled circle, the solid lines indicate the couplings $m_{1},m_{2},im_{3},im_{4}$ between nearest neighbor sites, and the dashed lines indicate the couplings $i\mu$ intra-site.

\subsection{Analytic results of the defective MZMs }
To acquire the correspondence of topological number $\mathcal{Z}$ and the number of edge stats for the finite size chain $\mathcal{C}_{\mathrm{finite}}$, we can investigate the analytic expression of edge states under the open boundary condition. We begin from the zero-mode eigenstates in the semi-infinite limit from the right or left boundary.
\begin{figure}[t]
\scalebox{0.34}{\includegraphics*[0.5in,0.0in][10.7in,5.8in]{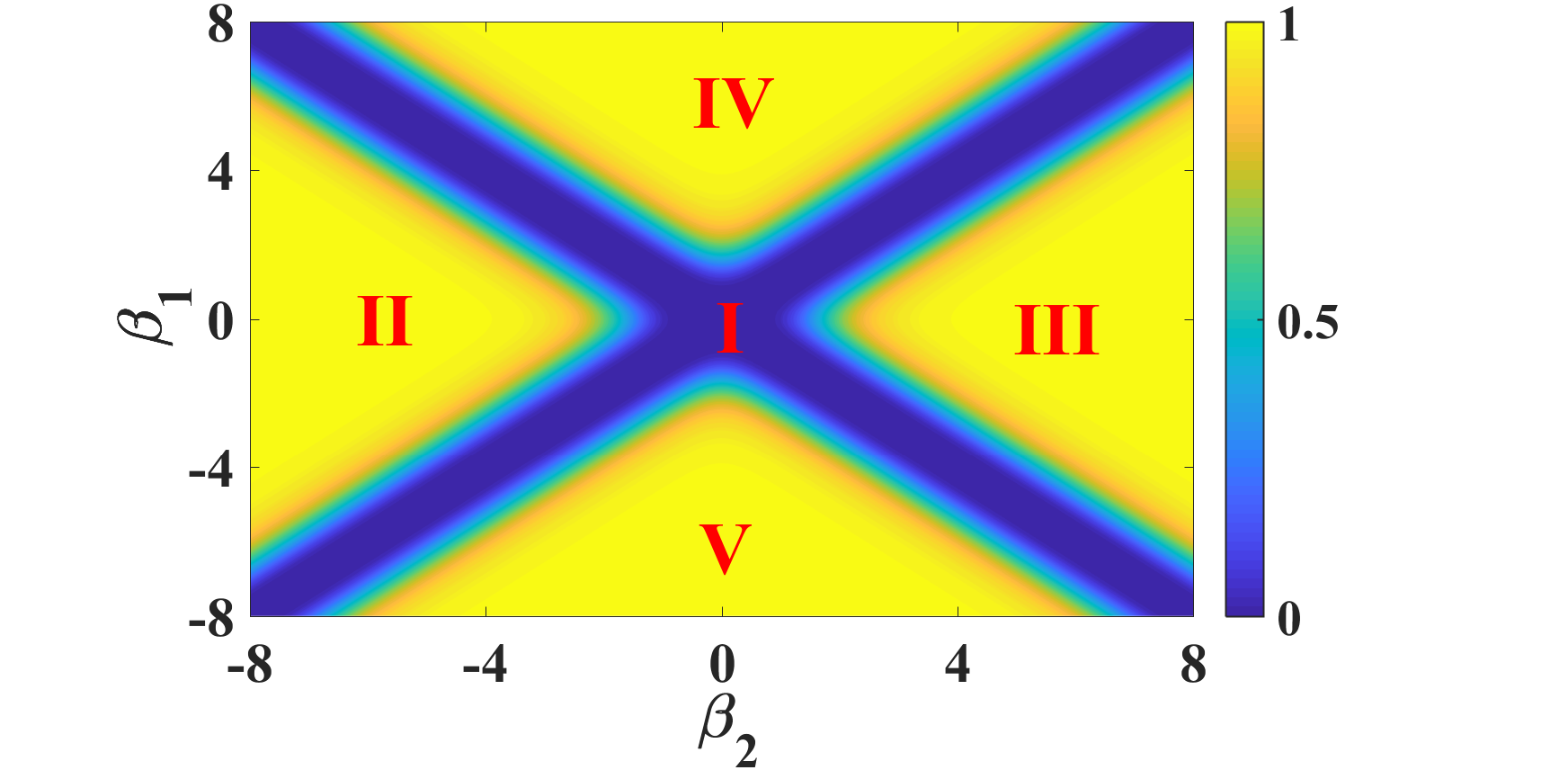}}
\caption{Similarity $\gamma _{\mathrm{spin}}$ of the two ground states in spin-representation with $L=4$, $t=\Delta_{0}=1$ and $\mu=0.1$. The blue area I represents $\gamma _{\mathrm{spin}}=0$ which means the two ground states are orthogonal, The yellow area II-V represent $\gamma _{\mathrm{spin}}=1$ which means the two ground states coalesce to one and act as exceptional points (EPs). }
 \label{PhaseDiagramSpin}
\end{figure}

In order to avoid confusion with the operators $c_{j},c_{j}^{\dagger}$ in $H_{\mathrm{NH}}$, we rewrite it's Hermitian counterpart $H_{\mathrm{cp}}$ in Eq.(\ref{Hcp}) by substituting the notation simply: $%
\{d_{j},d_{k}^{\dagger }\}\rightarrow \{\bar{c}_{j},\bar{c}_{k}^{\dagger
}\}$, i.e., $H_{\mathrm{cp}}=\bar{C}^{\dagger }h_{\mathrm{cp}}\bar{C}$, which is written as
\begin{equation}
H_{\mathrm{cp}}=-\sum_{j}\{(t\bar{c}_{j+1}^{\dagger }\bar{c}_{j}+\Delta _{0}\bar{c%
}_{j}^{\dagger }\bar{c}_{j+1}^{\dagger }+h.c.)-\mu (1-2\bar{n}_{j})\},
\label{HerCoun}
\end{equation}
this is just the 1D Hermitian Kitaev model. Therefore, we can investigate the non-Hermitian properties of $H_{\mathrm{NH}}$ based on its Hermitian counterpart $H_{\mathrm{cp}}$, with the definition of Majorana operators%
\begin{equation}
\bar{c}_{j}^{\dagger }=\frac{1}{2}(\bar{a}_{j}+i\bar{b}_{j}),\bar{c}_{j}=%
\frac{1}{2}(\bar{a}_{j}-i\bar{b}_{j})  \label{DiracMajTran}
\end{equation}%
we rewrite the Hamiltonian $H_{\mathrm{cp}}$ by Majorana operators,
\begin{equation}
H_{\mathrm{cp}}^{\mathrm{M}}=\frac{-i}{2}\sum_{j=1}^{L}\{(t-\Delta _{0})\bar{b}_{j+1}\bar{%
a}_{j}+(t+\Delta _{0})\bar{b}_{j}\bar{a}_{j+1}+2\mu \bar{a}_{j}\bar{b}%
_{j}\}
\end{equation}
The lattice schematic diagram of $H_{\mathrm{cp}}^{\mathrm{M}}$ is shown in Fig.\ref{chain}\textcolor[rgb]{0.00,0.00,1.00}{(c)}. As we know, in the Hermitian cases, two unpaired Majorana zero mode would locate at the right and left end of the Kitaev chain. Besides, for a finite size chain with $\mu\neq0$, the eigne energy of the two Majorana edge modes can split due to the coupling of them.

Next, we try to acquire the analytic expression for the edge states. Firstly, we consider a very long wire, which means the Majorana edges modes have zero energy; We
calculate their wave functions by using the Heisenberg equations of
motion
\begin{equation}
\lbrack H_{\mathrm{cp}}^{\mathrm{M}},\bar{a}_{m}]=0,[H_{\mathrm{cp}}^{\mathrm{M}},\bar{b}_{m}]=0,
\end{equation}%
where $\bar{a}_{m}$, $\bar{b}_{m}$ are the two Majorana operators at site n
which were denoted by $\bar{a}_{n,A},\bar{a}_{n,B}$ and $\bar{b}_{n,A},\bar{b%
}_{n,B}$ above. We obtain the difference equations for these operators as \cite{SemiFinit,SemiFinit2}:
\begin{eqnarray}
(t+\Delta _{0})\bar{b}_{m}+(t-\Delta _{0})\bar{b}_{m+1}+2\mu \bar{b}_{m} &=&0
\notag \\
(t-\Delta _{0})\bar{a}_{m}+(t+\Delta _{0})\bar{a}_{m+1}+2\mu \bar{a}_{m} &=&0
\end{eqnarray}%
for $2\leq m\leq 2L-1$, These difference equations can be solved exactly
by using $Z$-transform methods. We introducing a power series%
\begin{equation}
A(z)=\sum_{m=0}^{\infty }z^{-m}\bar{a}_{m}\equiv Z[\bar{a}_{m}]
\end{equation}%
where $z$ is a complex variable. The function $A(z)=Z[\bar{a}_{m}]$ is called
the $Z$-transform of $\bar{a}_{m}$. Taking the $Z$-transform of the above difference
equation and using properties such
\begin{equation}
Z[\bar{a}_{m-1}]=z^{-1}A(z),Z[\bar{a}_{m+1}]=zA(z)-z\bar{a}_{0}
\end{equation}%
$\bar{a}_{0}$ is a constant determined by boundary conditions, one can
obtain a closed form expression for the $Z$-transform $A(z)$, given by:%
\begin{equation}
A(z)=\frac{\bar{a}_{0}z^{2}}{z^{2}+\frac{\mu }{t-\Delta _{0}}z+\frac{%
t+\Delta _{0}}{t-\Delta _{0}}}
\end{equation}%
This $Z$-transform has a unique inverse, which is the exact solution to the
difference equation. Thus the obtained wave function is
\begin{subequations}
\begin{eqnarray}
\bar{a}_{m} &=&\bar{a}_{0}C^{m}\left[ \cos (\theta m)+\frac{1}{\tan \theta }%
\sin (\theta m)\right]  \label{aOC} \\
\bar{b}_{m} &=&\bar{b}_{0}C^{-m}\left[ \cos (\theta m)+\frac{1}{\tan \theta }%
\sin (\theta m)\right]  \label{bOC}
\end{eqnarray}%
\end{subequations}
where
\begin{equation}
C=\sqrt{\frac{t-\Delta _{0}}{t+\Delta _{0}}},\theta =\arctan \frac{\sqrt{%
t^{2}-\Delta _{0}^{2}-\mu ^{2}}}{\mu }  \label{Sita}
\end{equation}%

The entire model is equivalent to two coupled SSH like chains containing both the hopping parameters $i(\Delta _{0}-t)$ and $%
i(\Delta _{0}+t)$, as shown in Fig.\ref{chain}\textcolor[rgb]{0.00,0.00,1.00}{(c)}. Here, we only concentrate on the zero energy eigne
states, the excited states in complex and is not important here. If $%
t>\Delta _{0}>0,\mu \neq 0$, the system has two gap states, one can get
the analytical wave function of the zero energy mode (gap states) in
Majorana-representation. Here we express the zero mode of $H_{\mathrm{cp}}^{\mathrm{M}}$ in $\bar{a}$-sublattice and $\bar{b}$-sublattice as (i.e. the states in the left and right edges)
\begin{eqnarray}
\left\vert \psi \right\rangle _{a}^{L} &=&\sum_{m=0}^{L-1}\Lambda (m)\left\vert
m\right\rangle \otimes \left(
\begin{array}{c}
\bar{a} \\
0%
\end{array}%
\right) \nonumber \\
\left\vert \psi \right\rangle _{b}^{R} &=&\sum_{m=0}^{L-1}\Lambda (m)\left\vert
L-m\right\rangle \otimes \left(
\begin{array}{c}
0 \\
\bar{b}%
\end{array}%
\right)
\end{eqnarray}%
where we set $\bar{a}_{0,A}=\bar{b}_{N,B}=1$, $\Lambda (m)=C^{m}\Theta (m)$
\begin{equation}
\Theta (m)=\left[ \cos
(\theta m)+\frac{1}{\tan \theta }\sin (\theta m)\right] . \label{Sita}
\end{equation}%
When the region we considerate is $\mu
^{2}<4(t^{2}-\Delta _{0}^{2})$ predict an oscillatory
exponential decay of the coefficients: $e^{-n/2\zeta }$ for $\left\vert
\psi \right\rangle _{\bar{a}}$, where the decay length $\zeta$ is defined by
\begin{equation}
\zeta =1/\left\vert \ln \left( \frac{t-\Delta _{0}}{t+\Delta _{0}}\right)
\right\vert
\end{equation}%
So with the inverse similarity transformation based on Eq. (\ref{simTran}), Eq. (\ref{HerCoun}) and Eq. (\ref{DiracMajTran}), we exchange
the operator by the rule $(\bar{a}_{j},\bar{b}_{j})\rightarrow (\bar{c}%
_{j}^{\dagger },\bar{c}_{j})\rightarrow (d_{j}^{\dagger },d_{j})\rightarrow
(c_{j}^{\dagger },c_{j})$, and we get the anti-zero-energy modes (edge states) for
the initial non-Hermitian Hamiltonian $H_{\mathrm{NH}}$ in Nambu representation, as shown in Eq.(\ref{NambRep}). The right-vectors of the right/left localized edge states are expressed as
\begin{eqnarray}
\left\vert \psi _{\mathrm{NH}}^{\mathrm{R}} \right\rangle &=&\frac{i}{\sqrt{\mathcal{N}_{b}}}%
\sum_{m=1}^{L}\Lambda (L-m+1)[\Omega _{m}^{-1}\left\vert m\right\rangle -\Omega _{m}\left\vert L+m\right\rangle ],\nonumber \\
\left\vert \psi_{\mathrm{NH}}^{\mathrm{L}} \right\rangle &=&\frac{1}{\sqrt{\mathcal{N}_{a}}}%
\sum_{m=1}^{L}\Lambda (m)[\Omega _{m}^{-1}\left\vert m\right\rangle +\Omega _{m}\left\vert L+m\right\rangle ],
\label{Rvect}
\end{eqnarray}
where $L=2N$ is the length of the Majorana ladder, $\Omega _{\mathrm{m\in odd}}=e^{(\beta _{1}-\beta _{2})/2}$, $\Omega _{\mathrm{m\in even}}=e^{-(\beta _{1}+\beta _{2})/2}$.

\begin{figure}[t]
\includegraphics[clip,width=0.48\textwidth]{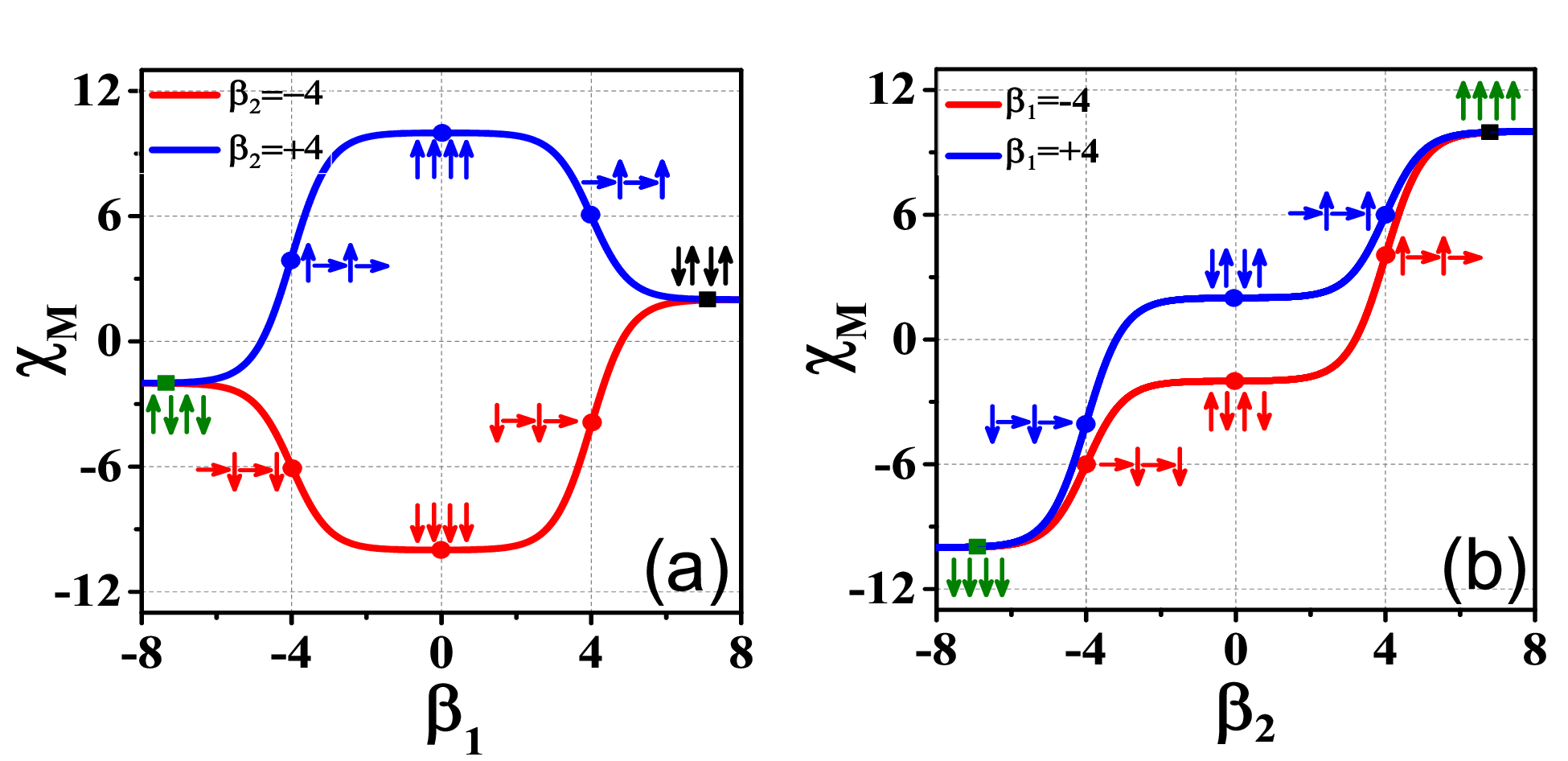}
\caption{The spin structure of the many-body ground state $\left\vert G\right\rangle _{\mathrm{NH}}^{1}$ with (a) $\beta_{2}=\pm4$ and (b) $\beta_{1}=\pm4$. Here, we set $L=4$, $t=\Delta_{0}=1$ and $\mu=0.1$. The non-Hermitian suppression effect, induced by the competition of $\beta_{1}$ and $\beta_{2}$, drives the spin flipping to opposite direction continuously without gap closing.}
 \label{Spinstructure}
\end{figure}
Similarly, we can get the left-vector $\left\langle \Phi
\right\vert $ of the zero mode edge states defined by $\left\langle \Phi
\right\vert H=\left\langle \Phi \right\vert E$ , or $H^{\dagger }\left\vert
\Phi \right\rangle =E^{\ast }\left\vert \Phi \right\rangle$. We can use the exchange $te^{-\beta _{1}}\leftrightarrow te^{\beta
_{1}}$, $\Delta _{0}e^{\beta _{2}}\leftrightarrow \Delta _{0}e^{-\beta _{2}}$, $\Omega \leftrightarrow \Omega ^{-1},$ for
convenience, the left-vector for the right/left localized edges states can be expressed as
\begin{eqnarray}
\left\vert \Phi _{\mathrm{NH}}^{\mathrm{R}} \right\rangle &=&\frac{i}{\sqrt{\mathcal{N}_{b}}}%
\sum_{m=1}^{L}\Lambda (L-m+1)[\Omega _{m}\left\vert m\right\rangle -\Omega _{m}^{-1}\left\vert L+m\right\rangle ],\nonumber \\
\left\vert \Phi_{\mathrm{NH}}^{\mathrm{L}} \right\rangle &=&\frac{1}{\sqrt{\mathcal{N}_{a}}}%
\sum_{m=1}^{L}\Lambda (m)[\Omega _{m}\left\vert m\right\rangle +\Omega _{m}^{-1}\left\vert L+m\right\rangle ],
\end{eqnarray}
where $\mathcal{N}_{a}=\mathcal{N}_{b}=\sum_{m=1}^{L}2%
\Lambda (m)^{2}\equiv\mathcal{N}$ is biorthogonal normalization coefficient.

Next, taking $(\left\vert \psi _{\mathrm{NH}}^{\mathrm{R}} \right\rangle,\left\vert \psi _{\mathrm{NH}}^{\mathrm{R}} \right\rangle )$ and $(\left\langle \Phi_{\mathrm{NH}}^{\mathrm{L}}\right\vert, \left\langle \Phi_{\mathrm{NH}}^{\mathrm{R}}\right\vert)$ as the basis states, we construct the \emph{effective Hamiltonian} of edge states for the finite-size Kitaev chain as
\begin{eqnarray}
\mathcal{H}_{\mathrm{edge}}=\left(
\begin{array}{cc}
h_{\mathrm{LL}} & h_{\mathrm{LR}} \\
h_{\mathrm{RL}} & h_{\mathrm{RR}}%
\end{array}%
\right),
\end{eqnarray}
where the elements of $\mathcal{H}_{\mathrm{edge}}$ are defined as
$h_{\mathrm{I,J}}=\left\langle \Phi ^{\mathrm{I}}_{\mathrm{NH}}|H_{\mathrm{NH}}|\psi ^{\mathrm{J}}_{\mathrm{NH}}\right\rangle$ and $\mathrm{I,J=L,R}$. We have $h_{\mathrm{LL}}=h_{\mathrm{RR}}=0,h_{\mathrm{RL}}=h_{\mathrm{LR}}^{\ast}=i\xi$, i.e., \begin{eqnarray}
\mathcal{H}_{\mathrm{edge}}=\left(
\begin{array}{cc}
0 & -i\xi \\
i\xi & 0%
\end{array}%
\right)=\xi\sigma_{y}
\end{eqnarray} where $\sigma_{y}$ denotes the Pauli matrices acting on the subspace of two edge states and $\xi$ is the coupling coefficient of them \cite{SemiFinit3}:
\begin{eqnarray}
\xi&=&\frac{\sqrt{t^{2}-\Delta _{0}^{2}}^{L+1}}{4\sin ^{3}\theta}\{2\sin(\theta L)\nonumber \\
&+&L\sin[\left( L+2\right) \theta ]
-L\sin \left[ \left( L+4\right) \theta \right]\}
\end{eqnarray} The energy of MZMs are $E_{\mathrm{edge}}^{\pm}=\pm \sqrt{|\xi|}$ and in the thermodynamic limit ($L\mapsto +\infty$) we have $E^{\pm}_{\mathrm{edge}}\mapsto 0$. Because the eigenstates of $\sigma _{y}$ is $(1,i)$ and $(1,-i)$, the eigenstates of $\mathcal{H}_{\mathrm{edge}}$ (i.e. the eigen wavefunction of edge states) are
\begin{eqnarray}
\left\vert \psi^{+} \right\rangle &=&\frac{1}{\sqrt{2\mathcal{N}}}
\left( \left\vert \psi _{\mathrm{NH}}^{\mathrm{L}}\right\rangle+i\left\vert \psi_{\mathrm{NH}}^{\mathrm{R}}
\right\rangle\right),\nonumber \\
\left\vert \psi^{-} \right\rangle &=&\frac{1}{\sqrt{2\mathcal{N}}}%
\left( \left\vert \psi_{\mathrm{NH}}^{\mathrm{L}} \right\rangle-i\left\vert \psi_{\mathrm{NH}}^{\mathrm{R}}
\right\rangle\right).
\end{eqnarray}
So far, we get the analytic solution of the non-Hermitian Majorana zero modes.

\subsection{Number-anomalous bulk-boundary correspondence}
Then, to describe the localization and the orthogonality of the two edge states, i.e., to investigate the bulk-boundary correspondence of the NH system, we define the similarity of the two edge states as
\begin{equation}
\gamma _{\mathrm{M}}\equiv\left\langle \psi^{-}|\psi^{+}\right\rangle =\left( e^{-\beta
_{1}}-e^{\beta _{1}}\right) \left( e^{\beta _{2}}-e^{-\beta _{2}}\right) \kappa,
\label{gamaM}
\end{equation}where $\kappa =\frac{1}{2\mathcal{N}}\sum_{n=0}^{L-1}(-1)^{n}C^{2n}\Theta
_{n}^{2}$ is a nonzero value independent of $\beta_{1}$ and $\beta_{2}$. When $\beta_{1,2}\neq 0$ we have $\gamma_{\mathrm{M}}\neq0$, which means the two edge states $\left\vert\psi^{\pm}\right\rangle$ are not orthogonal and no longer located at two ends of the chain respectively. The distributions of similarity $\gamma _{\mathrm{M}}(\beta_{1},\beta_{2})$ and corresponding eigen wavefunction are calculated numerically and are shown in Fig.\ref{PhaseDiagramMajorana}. We can see that, $\gamma _{\mathrm{M}}$ tends to be 1 when the NH strength $\beta_{1}$ and $\beta_{2}$ are away from 0, this means the two MZMs become one gradually and evolve to the exceptional points (EPs) eventually. As a result, the typical BBC is broken and the MZM local only at the left or right end of the system. For example, when $\beta_{1}=\beta_{2}=3$ (point P5 in Fig.\ref{PhaseDiagramMajorana}\textcolor[rgb]{0.00,0.00,1.00}{(a)}), $\gamma _{\mathrm{M}}=1$ and the Majorana zero mode only locate at left end of the chain (shown in (b5) of Fig.\ref{PhaseDiagramMajorana}\textcolor[rgb]{0.00,0.00,1.00}{(b)}).

Therefore, the number-anomalous BBC can be defined as \cite{kou2020}
\begin{eqnarray}
\mathcal{C}_{\mathrm{finite}}= 2-\gamma _{\mathrm{M}}.
\end{eqnarray}
In the Hermitian case ($\beta_{1}=\beta_{2}=0$), we have $\gamma _{\mathrm{M}}=0$, the number of MZMs is $\mathcal{C}_{\mathrm{finite}}=2$ and the typical BBC is acquired. While, in the NH cases ($\beta_{1,2}\neq0$), we have $0<\gamma _{\mathrm{M}}\leq1$, the MZMs becomes defective and we have $1\leq\mathcal{C}_{\mathrm{finite}}\leq2$. A special case is $\mathcal{C}_{\mathrm{finite}}=1$, which indicates the existence of a singular MZM. This phenomenon didn't occur in previous studies, neither in Hermitian nor in NH SC systems.

There are two reasons for this observation: the breakdown of sublattice symmetry induced by $\beta_{1}$ and the breakdown of the particle-hole symmetry induced by $\beta_{2}$. As shown in the Fig.\ref{chain}\textcolor[rgb]{0.00,0.00,1.00}{(a)}, the imbalanced particle hopping lead to the breakdown of sublattice symmetry, which can suppress the particles located at A-sublattice ($\beta_{1}<0$) or B-sublattice ($\beta_{1}>0$). Meanwhile, the imbalanced paring amplitudes lead to the breakdown of the particle-hole symmetry, which make the Majorana quasi-particle behave particle-like or hole-like. In a word, the defective Majorana zero modes are induced by the breakdown of sublattice symmetry and particle-hole symmetry.

\begin{figure}[t]
\includegraphics[clip,width=0.45\textwidth]{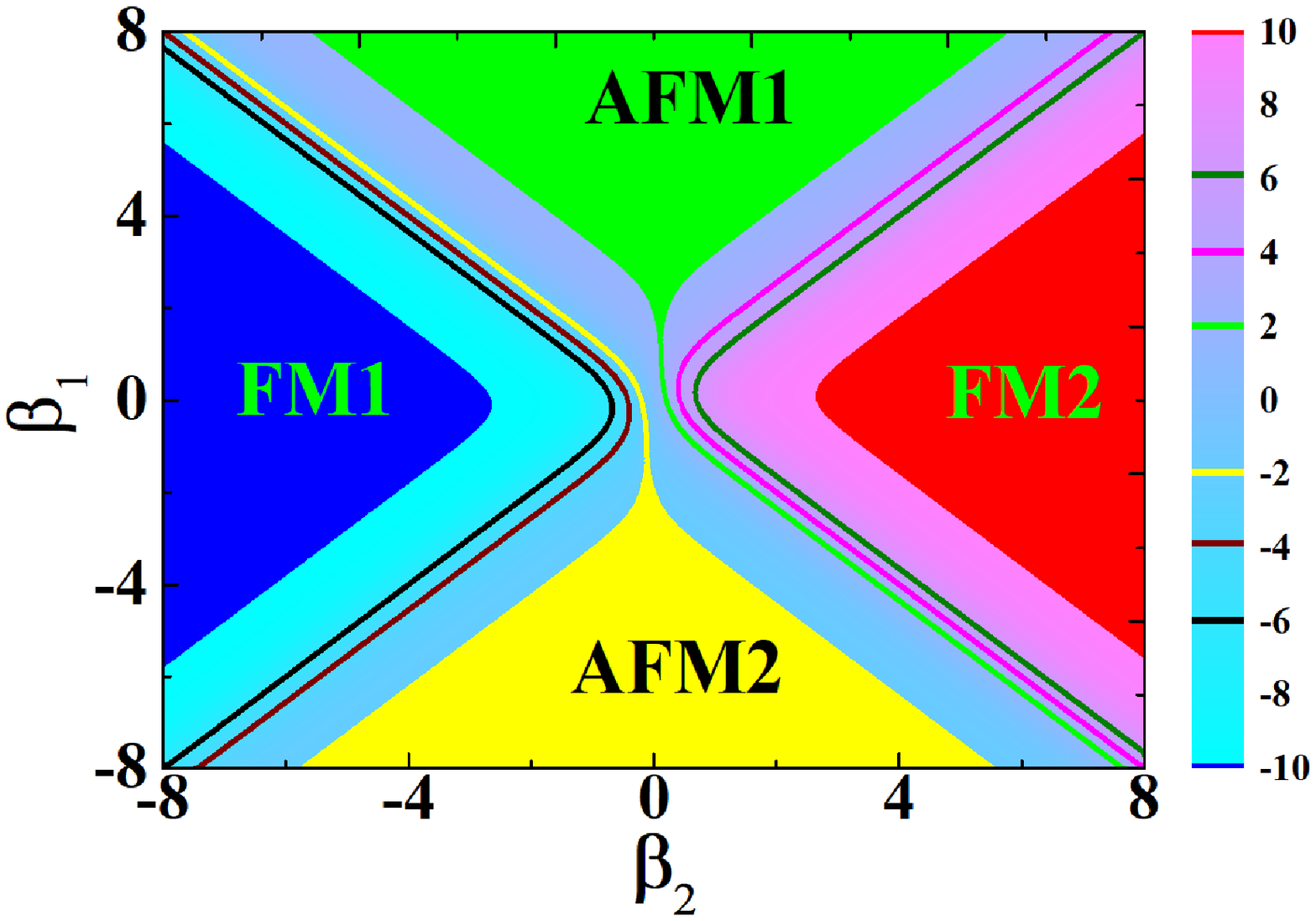}
\caption{Global magnetic phase diagram of the many-body ground states. The variations of $\chi _{\mathrm{M}}$ induced by non-Hermitian suppression effect are marked by different colors: FM phase (blue and red area), AFM phase (green and yellow area) and so on. Here, we set $L=4$, $t=\Delta_{0}=1$ and $\mu=0.1$.}
\label{MagPhase}
\end{figure}

\section{Many-body correspondence of the defective MZMs}\label{ManyBody}

Benefit from the mapping between Kitaev chain and spin chain, one can simulate the MZMs in the Ising language via a Jordan-Wigner transformation  \cite{JWTran1,JWTran2}, because the Ising model is relatively easy to implement. For example, MZMs have attracted much attention due to their potential application in topological quantum computations, the spin chain can be used to simulate the braiding of these MZMs, which corresponding to the topological quantum gates \cite{Guo2016}. Therefore, it is necessary to explore the many-body correspondence of the defective MZMs and explain the NH effects on the spin-representation.
\begin{table*}[t]
\begin{tabular}{|c|c|c|c|c|c|}
\hline
$\ \ \ \ \ \ \ \ \ \ \ \ \ \ \ \ \ \ \ \ \ $ & $\ \ \ \left\vert
G\right\rangle _{\mathrm{NH}}^{1}\ \ \ $ & $\ \ \ \left\vert G\right\rangle
_{\mathrm{NH}}^{2}\ \ \ $ & $\ \ \ \ \ \ \ \ \chi _{\mathrm{M}}\ \ \ \ \ \ \ \ \ $ & $\gamma
_{\mathrm{spin}}$ & Phase \\ \hline
$\beta _{1}=0,\beta _{2}\rightarrow +\infty $ & $\left\vert \uparrow
\uparrow \uparrow \uparrow \right\rangle $ & $\left\vert \uparrow \uparrow
\uparrow \uparrow \right\rangle $ & $L(L+1)/2$ & $1$ & FM2 \\ \hline
$\beta _{1}=0,\beta _{2}\rightarrow -\infty $ & $\left\vert \downarrow
\downarrow \downarrow \downarrow \right\rangle $ & $\left\vert \downarrow
\downarrow \downarrow \downarrow \right\rangle $ & $-L(L+1)/2$ & $1$ & FM1 \\
\hline
$\beta _{2}=0,\beta _{1}\rightarrow +\infty $ & $\left\vert \downarrow
\uparrow \downarrow \uparrow \right\rangle $ & $\left\vert \downarrow
\uparrow \downarrow \uparrow \right\rangle $ & $L/2$ & $1$ & AFM1 \\ \hline
$\beta _{2}=0,\beta _{1}\rightarrow -\infty $ & $\left\vert \uparrow
\downarrow \uparrow \downarrow \right\rangle $ & $\left\vert \uparrow
\downarrow \uparrow \downarrow \right\rangle $ & $-L/2$ & $1$ & AFM2 \\ \hline
$\beta _{1}=\beta _{2}>0$ & $\left\vert \rightarrow \uparrow
\rightarrow \uparrow \right\rangle $ & $\left\vert \leftarrow
\uparrow \leftarrow \uparrow \right\rangle $ & $L(L+2)/4$ & $0$ & Even-FM2
\\ \hline
$\beta _{1}=\beta _{2}<0$ & $\left\vert \rightarrow \downarrow
\rightarrow \downarrow \right\rangle $ & $\left\vert \leftarrow
\downarrow \leftarrow \downarrow \right\rangle $ & $-L(L+2)/4$ & $0$ &
Even-FM1 \\ \hline
$\beta _{1}=-\beta _{2}<0$ & $\left\vert \uparrow \rightarrow \uparrow
\rightarrow \right\rangle $ & $\left\vert \uparrow \leftarrow
\uparrow \leftarrow \right\rangle $ & $L^{2}/4$ & $0$ & Odd-FM2 \\ \hline
$\beta _{1}=-\beta _{2}>0$ & $\left\vert \downarrow \rightarrow
\downarrow \rightarrow \right\rangle $ & $\left\vert \downarrow
\leftarrow \downarrow \leftarrow \right\rangle $ & $-L^{2}/4$ & $0$
& Odd-FM1 \\ \hline
\end{tabular}
\caption{The non-Hermitian effects on the two ground states in different limit cases (first collum). The spin configurations, magnetic factor $\chi _{\mathrm{M}}$, similarity rate $\gamma_{\mathrm{spin}}$ and magnetic phase are shown in collum 2-6, respectively. }
\label{PhaseDiagramTable}
\end{table*}

\subsection{NH spin model corresponding to the NH Kitaev model}
In the Hermitian case, it has been noted that the Kitaev model can be mapped to the Ising model via the Jordan-Wigner transformation, and the two MZMs can also be mapped to the two degenerate ground states of the Ising model. Similarly, we can transform the NH Kitaev model to NH Ising model in this way. The Hamiltonian $H_{\mathrm{NH}}$ can be written in spin-representation as
\begin{eqnarray}
H_{\mathrm{NH}}^{\mathrm{spin}} &=&-\frac{1}{4}\sum_{j}\{J_{1}\sigma _{j}^{x}\sigma
_{j+1}^{x}+J_{2}\sigma _{j}^{y}\sigma _{j+1}^{y}\notag \\
&&+iJ_{3}\sigma _{j}^{x}\sigma _{j+1}^{y}+iJ_{4}\sigma
_{j}^{y}\sigma _{j+1}^{x}+4\mu \sigma _{j}^{z}\},
\end{eqnarray}
where
\begin{eqnarray}
J_{1}&=&t\cosh \beta _{1}+\Delta _{0}\cosh \beta _{2},\nonumber\\
J_{2}&=&t\cosh \beta _{1}-\Delta _{0}\cosh \beta _{2},\nonumber\\
J_{3}&=&\epsilon t\sinh\beta _{1}+\Delta _{0}\sinh \beta _{2},\nonumber\\
J_{4}&=&-\epsilon t\sinh \beta _{1}+\Delta _{0}\sinh \beta _{2}.
\end{eqnarray}
And $H_{\mathrm{NH}}^{\mathrm{spin}}$ can also be transformed to a Hermitian Hamiltonian by the similarity transformation
\begin{eqnarray}
H_{\mathrm{cp}}^{\mathrm{spin}}=UH_{\mathrm{NH}}^{\mathrm{spin}}U^{-1},
\label{spinTran}
\end{eqnarray}
where the similarity transformation operator can be expressed as
$U={\prod}_{j}\otimes U_{j,A}\otimes U_{j,B},$ where
\begin{eqnarray}
U_{j,A}=diag\{1,\mathrm{exp}[(\beta _{2}-\beta _{1})/2]\},\nonumber\\
U_{j,B}=diag\{1,\mathrm{exp}[(\beta _{2}+\beta _{1})/2]\}.
\end{eqnarray}
Therefore, the Hermitian counterpart of $H_{\mathrm{NH}}$ in spin-representation can be expressed as \begin{eqnarray}
H_{\mathrm{cp}}^{\mathrm{spin}}=-\sum_{j=1}^{L-1}\{J_{x}\sigma _{j}^{x}\sigma
_{j+1}^{x}+J_{y}\sigma _{j}^{y}\sigma _{j+1}^{y}-2\mu \sigma
_{j}^{z}\},
\end{eqnarray}
where $J_{x/y}=t\pm\Delta _{0}$ and this is just the quantum XY spin
chain and its extensions have been studied from
many different perspectives.

\subsection{Similarity of the two degenerate ground states}

To characterize the properties of ground states, we calculate
the magnetic factor when the system is in its
ground state. Firstly, for simplicity and typicality, we consider the limit case $\mu =0$, $t=\Delta _{0}$ where $H_{\mathrm{cp}}^{\mathrm{spin}}$ is exactly the Ising model without transverse field, the ground states of $H_{\mathrm{cp}}^{\mathrm{spin}}$ are described as
\begin{eqnarray}
\left\vert G\right\rangle _{\mathrm{cp}}^{1} =\left\vert \rightarrow
\rightarrow \cdots \rightarrow \right\rangle,%\nonumber\\
\left\vert G\right\rangle _{\mathrm{cp}}^{2} =\left\vert \leftarrow
\leftarrow \cdots \leftarrow \right\rangle.
\end{eqnarray}
So the ground states of initial NH spin chain can be
obtained completely under the inverse similarity transformation:
\begin{eqnarray}
\left\vert G\right\rangle _{\mathrm{NH}}^{1}=U^{-1}\left\vert G\right\rangle
_{\mathrm{cp}}^{1},
\left\vert G\right\rangle _{\mathrm{NH}}^{2}=U^{-1}\left\vert
G\right\rangle _{\mathrm{cp}}^{2}.
\end{eqnarray}
In fact, $U_{i}$ act as a transverse field inner each site $i$. When we adjust $\beta _{1}$ and $\beta _{2}$, the two ground states may construct different spin structure in the spin chain. Then, we define the similarity of the two ground states as
\begin{equation}
\gamma _{\mathrm{spin}}(\beta _{1},\beta _{2})={}_{\mathrm{NH}}^{~~~2}\left\langle G|G\right\rangle
^1_{\mathrm{NH}},
\end{equation}%
After a tedious calculation, we can obtain
\begin{equation}
\gamma _{\mathrm{spin}}(\beta _{1},\beta _{2})=(-\tanh[\frac{(\beta _{1}+\beta _{2})}{2}]\tanh[\frac{(\beta _{1}-\beta _{2})}{2}])^{\frac{L}{2}}
\end{equation}
Therefore, $\gamma _{\mathrm{spin}}$ will change with $\beta _{1}$ and $\beta _{2}$. It is obvious that $\left\vert G\right\rangle _{\mathrm{NH}}^{1}$ and $\left\vert G\right\rangle _{\mathrm{NH}}^{2}$ are not always orthogonal. We study this phenomena in some limiting case: (1) when $\beta_1=0$, we have $\gamma _{\mathrm{spin}}(0,\beta _{2})=[\tanh(\frac{\beta _{2}}{2})]^{L}$; (2) When $\beta_2=0$, we have $\gamma _{\mathrm{spin}}(\beta _{1},0)=[-\tanh(\frac{\beta _{1}}{2})]^{L}$;  (3) when $\beta _{1}=\beta _{2}$, we have $\gamma _{\mathrm{spin}}(\beta _{1}=\beta _{2})=0$. Indeed, for spin systems with finite size, two ground states will coalesce when $\beta _{1}\mapsto 0$(or $\beta _{2}\mapsto 0$). While, two ground states can't coalesce in the thermodynamic limit $L\mapsto\infty$, due to $\tanh(\beta_{1/2})<1$.

Now, we take 4-spin systems as an example and show the coalescence phase diagram in Fig.(\ref{PhaseDiagramSpin}). We can see that $\gamma _{\mathrm{spin}}\mapsto0$ when the NH strengths are near $\beta _{1}=\beta _{2}$ region, and $\gamma _{\mathrm{spin}}\mapsto1$ when $\beta _{1},\beta _{2}$ are large enough which means the wave functions of the two degenerate ground states coalesce. This is very different from the similarity of MZM in Majorana representation shown in Fig.(\ref{PhaseDiagramMajorana}), where $\gamma _{\mathrm{M}}\rightarrow0$ in the region near $\beta _{1}=0$, or $\beta _{2}=0$.

\subsection{Phase crossover without gap closing}

An important question is why the coalescing phase diagram of
MZMs shown in Fig.(\ref{PhaseDiagramMajorana}) is totally different from the coalescing phase diagram of
spin ground states shown in Fig.(\ref{PhaseDiagramSpin}), i.e.
\begin{equation}
\gamma _{\mathrm{M}}(\beta _{1},\beta _{2})\neq \gamma
_{\mathrm{spin}}(\beta _{1},\beta _{2}).
 \end{equation}
The key point is the correspondence of  single-particle and the many-body systems. We know that the relation of MZMs and the ground states of spin system is
\begin{equation}
\left\vert G\right\rangle _{\mathrm{NH}}^{1/2}=\hat{\psi}^{\mathrm{L/R}}_{\mathrm{NH}}\left\vert F\right\rangle,
\end{equation}
where $\left\vert F\right\rangle$ is the many-body vacuum state in spin-representation, and it is also the many-body quantum state with occupied single particle states for $E<0$ and empty single particle states for $E\geq0$. Therefore, the single-body wave function $\left\vert\psi^{\mathrm{L/R}}_{\mathrm{NH}}\right\rangle$ shown in Eq.\ref{Rvect} can't describe the many-body ground states absolutely, we must take $\left\vert F\right\rangle$ into account. Indeed, the NH terms perturb the vacuum background also, even they do not change the energy of the states.

To give a quantitative description for the spin configurations of the two ground states in the spin system, we define a magnetic factor as
\begin{equation}
\chi _{\mathrm{M}}(\beta _{1},\beta _{2})\equiv\sum_{n=1}^{L}{~}_{\mathrm{NH}}^{~~~1}\left\langle G\right\vert\negmedspace
n\sigma _{n}^{z}\left\vert G\right\rangle _{\mathrm{NH}}^{1},
\end{equation}%
where a weighted sum of the spin of the n-th lattice is introduced. The introduce of the lattice number "n" can distinguish each site, so the magnetic factor $\chi _{\mathrm{M}}(\beta _{1},\beta _{2})$ contains all spin information in each site. Then, we take the lattice size $L=4$ as an example, and give the magnetic susceptibility and spin configurations of the many-body ground states $\left\vert G\right\rangle _{\mathrm{NH}}^{1}$ for typical cases $\beta_{2}=\pm 4$ in Fig.(\ref{Spinstructure}a) and $\beta_{1}=\pm 4$ in Fig.(\ref{Spinstructure}b), where $L=4, t=\Delta_{0}=1, \mu=0.1$. Besides, we summarize the phase for different limit cases in Table \ref{PhaseDiagramTable}. We can see that the competition of $\beta_{1}$ and $\beta_{2}$ can drive the spin flipping to opposite direction continuously, this is the non-Hermitian suppression effect. Moreover the global magnetic order phase diagram of the many-body ground states are shown in Fig.\ref{MagPhase}, which can be divided into five part corresponding to the
coalesce phase diagram: region II and III are ferromagnetic (FM1,FM2), region IV
and V are antiferromagnetic (AF1, AF2), while region I are the transition
area of those four phases.

It should be emphasized that the gap remains opened during the phase crossover. The non-Hermitian Hamiltonian $H_{\mathrm{NH}}^{\mathrm{spin}}$ can be transformed to a Hermitian Hamiltonian by the similarity transformation as shown in Eq.(\ref{spinTran}). Because the similarity transformation can not change the energy spectrum, the gap between the ground states and the first exited state don't vary with $\beta_{1/2}$, the phase diagram is obtained without gap closing.

\section{Conclusion}\label{Summary}
We investigate the non-Hermitian effects on the Kitaev chain, whose hopping and superconductor paring strength are both imbalanced. The biorthogonal $\mathcal{Z}_{2}$ topological invariant and Majorana edge states are given analytically, these two imbalanced NH terms can induce defective Majorana edge states, which means one of the two localized edge states will disappear due to the NH suppression effect. As a result, the typical bulk-boundary correspondence is broken down. Besides, the defective edge states are mapped to the ground states of the non-Hermitian transverse field Ising model. With the definition of state-similarity and magnetic factor for the two ground states, the spin structure and the global phase diagrams are given, the FM-AFM crossover without gap closing is revealed. The novel non-Hermitian effects may provide a way to investigate MZMs and topological physics.

\section{Acknowledgments}
This work is supported by NSFC Grant No. 11674026, 11974053, 1217040237, 61835013, National Key R$\&$D Program of China under grants No.
2016YFA0301500, Strategic Priority Research Program of the Chinese Academy of
Sciences under grants Nos. XDB01020300, XDB21030300.


\begin{thebibliography}{99}

%%%%%%%%%%%%%%%%%%%%%%%%%%%%%%%%%
\bibitem {kitaev2001}A. Y. Kitaev, Phys. Usp. \textbf{44}, 131 (2001).
%Unpaired Majorana Fermions in Quantum Wires.
\bibitem {Fu2008}L. Fu, and C. L. Kane, Phys. Rev. Lett. \textbf{100}, 096407
(2008).
%Superconducting proximity effect and Majorana fermions at the surface of a topological insulator.

\bibitem {Mourik2012}V. Mourik, K. Zuo1, S. M. Frolov, S. R. Plissard, E. P. A. M. Bakkers, and L. P. Kouwenhoven, Science \textbf{336}, 1003 (2012).
%Signatures of Majorana fermions in hybrid superconductor-semiconductor nanowire devices.


\bibitem {Deng2012}M. T. Deng, C. L. Yu, G. Y. Huang, M. Larsson, P. Caroff, and H. Q. Xu
, Nano. Lett. \textbf{12}, 6414 (2012).
%Anomalous zero-bias conductance peak in a Nb-InSb nanowire-Nb hybride device.


\bibitem {Rokhinson2012}L. P. Rokhinson, X. Liu, and J. K. Furdyna, Nat. Phys.
\textbf{8}, 795 (2012).
%The fractional a.c. Josephson effect in a semiconductor-superconductor nanowire as a signature of Majorana particles.


\bibitem {Alicea2012}J. Alicea, Rep. Prog. Phys. \textbf{75}, 076501 (2012)
%New directions in the pursuit of Majorana fermions in solid state systems.


\bibitem {Mebrahtu2013}H. Mebrahtu, I. Borzenets, H. Zheng, Y. Bomze, A. I. Smirnov, S. Florens, H. U. Baranger, and G. Finkelstein, Nat. Phys. \textbf{9}, 732
(2013).
%Observation of Majorana quantum critical behaviour in a resonant level coupled to a dissipative environment.


\bibitem {Nadj-Perge2014}S. Nadjperge, I. K. Drozdov, J. Li, H. Chen, S. Jeon, J. Seo, A. H. Macdonald, B. A. Bernevig, and A. Yazdani, Science \textbf{346}, 602
(2014).
%Observation of Majorana fermions in ferromagnetic atomic chains on a superconductor.


\bibitem {Lee2014}E. J. H. Lee, X. Jiang, M. Houzet, R. Aguado, C. M. Lieber, and S. D. Franceschi, Nat. Nano \textbf{9}, 79 (2014).
%Spin-resolved Andreev levels and parity crossings in hybrid superconductor-semiconductor nanostructures.


\bibitem {read2000}N. Read, and D. Green, Phys. Rev. B \textbf{61}, 10267 (2000).
%Paired states of fermions in two dimensions with breaking of parity and time-reversal symmetries and the fractional quantum Hall effect.

\bibitem {Ivanov2001}D. A. Ivanov, Phys. Rev. Lett. \textbf{86}, 268 (2001).
%Non-Abelian statistics of half-quantum vortices in p-wave superconductors.

\bibitem {Sarma2006}S. DasSarma, C. Nayak, and S. Tewari, Phys. Rev. B
\textbf{73}, 220502(R) (2006).
%Proposal to stabilize and detect half-quantum vortices in strontium ruthenate thin films:Non-Abelian braiding statistics of vortices in a px+ipy superconductor
\bibitem {Stern2010}A. Stern, Nature (London) \textbf{464}, 187 (2010).
%Non-Abelian states of matter.

\bibitem {Tewari2007}B. Lian, X. Q. Sun, A. Vaezi, X. L. Qi, and S. C. Zhang, Proc. Natl. Acad. Sci. U.S.A. \textbf{115}, 10938 (2018).
%Topological quantum computation based on chiral Majorana fermions.

\bibitem {Nayak2008}C. Nayak, S. H. Simon, A. Stern, M. Freedman, and S. DasSarma, Rev. Mod. Phys. \textbf{80}, 1083 (2008).
%Non-Abelian anyons and topological quantum computation.






\bibitem {Sau2010}Jay D. Sau, Roman M. Lutchyn, Sumanta Tewari, and S. Das Sarma, Phys. Rev. Lett. \textbf{104}, 040502 (2010).

%Jay D. Sau, Roman M. Lutchyn, Sumanta Tewari, and S. Das Sarma Generic new platform for topological quantum computation using semiconductor heterostructures.

\bibitem {Alicea2011}J. Alicea, Y. Oreg, G. Refael, F. von Oppen, and M. P. A. Fisher, Nat. Phys. \textbf{7}, 412 (2011).
%Non-Abelian statistics and topological quantum information processing in 1D wire networks

\bibitem {Ashida2020}Y. Ashida, Z. Gong, and M. Ueda, Adv. Phys. \textbf{69}, 249 (2020).
%%%%Review Non-Hermitian Physics
\bibitem{Subsystem} I. Rotter and J. P. Bird, Rep. Prog. Phys. \textbf{78}, 114001 (2015).

\bibitem {EPReview2021}E. J. Bergholtz, J.C. Budich, and F. K. Kunst, Rev. Mod. Phys. \textbf{93}, 015005 (2021)
    %Exceptional topology of non-Hermitian systems


\bibitem {Yin2018}C. Yin, H. Jiang, L. Li, R. L\"{u}, and S. Chen, Phys. Rev.
A \textbf{97}, 052115 (2018).
%(Geometrical meaning of winding number and its characterization of topological phases in one-dimensional chiral non-Hermitian systems)

\bibitem {Ghatak2019}A. Ghatak and T. Das, J. Phys.: Condens. Matter
\textbf{31}, 263001 (2019).
%(New topological invariants in nonHermitian systems)

\bibitem {Rudner2009}M. S. Rudner and L. S. Levitov, Phys. Rev. Lett.
\textbf{102}, 065703 (2009).
%(Topological transition in a non-hermitian quantum walk)


\bibitem {Esaki2011}K. Esaki, M. Sato, K. Hasebe, and M. Kohmoto, Phys. Rev. B
\textbf{84}, 205128 (2011).
%(Edge states and topological phases in non-Hermitian systems)


\bibitem {Hu2011}Y. C. Hu and T. L. Hughes, Phys. Rev. B \textbf{84}, 153101
(2011).
%(Absence of topological insulator phases in nonHermitian -symmetric Hamiltonians)

\bibitem {Shen2018}H. Shen, B. Zhen, and L. Fu, Phys. Rev. Lett. \textbf{120},
146402 (2018).
%(Topological band theory for non-Hermitian Hamiltonians)

\bibitem {Lieu2018}S. Lieu, Phys. Rev. B \textbf{97}, 045106 (2018).
%(Topological phases in the non-Hermitian Su-Schrieffer-Heeger model)

\bibitem {Gong2018}Z. Gong, Y. Ashida, K. Kawabata, K. Takasan, S.
Higashikawa, and M. Ueda, Phys. Rev. X \textbf{8}, 031079 (2018).
%(Topological phases of non-Hermitian systems)

\bibitem {chen-class2019}C. H. Liu, H. Jiang, S. Chen, Phys. Rev. B
\textbf{99}, 125103 (2019).
%(Topological classification of non-Hermitian systems with reflection symmetry)

\bibitem {Jiang2018}H. Jiang, C. Yang, and S. Chen, Phys. Rev. A \textbf{98},
052116 (2018).
%(Topological invariants and phase diagrams for one-dimensional two-band non-Hermitian systems without chiral symmetry)

\bibitem {38-1}K. Kawabata, K. Shiozaki, M. Ueda, and M. Sato, Phys. Rev. X
\textbf{9}, 041015 (2019).
%(Symmetry and topology in non-Hermitian physics)

\bibitem {38}H. Zhou and J. Y. Lee, Phys. Rev. B \textbf{99}, 235112 (2019).
%(Periodic table for topological bands with non-Hermitian symmetries)

\bibitem {Kunst2019}F. K. Kunst and V. Dwivedi, Phys. Rev. B \textbf{99},
245116 (2019).
%(Non-Hermitian systems and topology: A transfer-matrix perspective)

\bibitem {Leykam2017}D. Leykam, K. Y. Bliokh, C. Huang, Y. D. Chong, and F.
Nori, Phys. Rev. Lett. \textbf{118}, 040401 (2017).
%(Edge modes, degeneracies, and topological numbers in non-Hermitian systems)

\bibitem {Lee2016}T. E. Lee, Phys. Rev. Lett. \textbf{116}, 133903 (2016).
%(Anomalous edge state in a non-Hermitian lattice)

\bibitem {KawabataUeda2018}K. Kawabata, K. Shiozaki, and M. Ueda, Phys. Rev. B
\textbf{98}, 165148 (2018).
%(Anomalous helical edge states in a non-Hermitian Chern insulator)

 \bibitem{kou2020} X. R. Wang, C. X. Guo, and S. P. Kou, Phys. Rev. B \textbf{101}, 121116(R) (2020)

\bibitem {Yao2018}S. Yao, and Z. Wang, Phys. Rev. Lett. \textbf{121}, 086803
(2018).
%(Edge states and topological invariants of non-Hermitian systems)


\bibitem {YaoWang2018}S. Yao, F. Song, and Z. Wang, Phys. Rev. Lett.
\textbf{121}, 136802 (2018).
%(Non-hermitian chern bands)Phys. Rev. Lett.\textbf{121}, 136802 (2018).Phys. Rev. Lett. \textbf{121}, 086803 (2018).

\bibitem {Deng2019}T. S. Deng and W. Yi, Phys. Rev. B \textbf{100}, 035102,
(2019).
%(Non-Bloch topological invariants in a non-Hermitian domain-wall system)(3,7)


\bibitem {SongWang2019}F. Song, S. Yao, and Z. Wang, Phys. Rev. Lett.
\textbf{123}, 170401 (2019).
%(Non-Hermitian skin effect and chiral damping in open quantum systems)(4,10)


\bibitem {Longhi2019}S. Longhi, Phys. Rev. Research \textbf{1}, 023013
(2019).
%(Probing non-Hermitian skin effect and non-Bloch phase transitions)


\bibitem {Xiong2018}Y. Xiong, J. Phys. Commun. \textbf{2}, 035043 (2018).
%(Why does bulk boundary correspondence fail in some non-hermitian
%topological models)


\bibitem {Kunst2018}F. K. Kunst, E. Edvardsson, J. C. Budich, and E. J.
Bergholtz, Phys. Rev. Lett. \textbf{121}, 026808 (2018).
%(Biorthogonal bulk-boundary correspondence in non-Hermitian systems)

\bibitem {Jin2019} S. Lin, L. Jin, and Z. Song, Phys. Rev. B \textbf{99}, 165148 (2019);
K. L. Zhang, H. C. Wu, L. Jin, and Z. Song, Phys. Rev. B \textbf{100}, 045141
(2019).
%(Bulk-boundary correspondence in a non-Hermitian system in one dimension with chiral inversion symmetry)
%(Symmetry protected topological phases characterized by isolated exceptional points)
%(Topological Phase Transition Independent of System Non-Hermiticity)

\bibitem {Lee2019}C. H. Lee and R. Thomale, Phys. Rev. B \textbf{99},
201103(R) (2019).
%(Anatomy of skin modes and topology in non-Hermitian systems)

\bibitem {Herviou2019}L. Herviou, J. H. Bardarson, and N. Regnault, Phys. Rev.
A \textbf{99,} 052118 (2019).
%(Defining a bulk-edge correspondence for non-Hermitian Hamiltonians via singular-value decomposition)

\bibitem {Yokomizo2019}K. Yokomizo and S. Murakami, Phys. Rev. Lett.
\textbf{123}, 066404 (2019).
%(non-Bloch Band Theory for Non-Hermitian Systems)

\bibitem {Zeuner2015}J. M. Zeuner, M. C. Rechtsman, Y. Plotnik, Y. Lumer, S.
Nolte, M. S. Rudner, M. Segev, and A. Szameit, Phys. Rev. Lett. \textbf{115},
040402 (2015).
%Shiyan(Observation of a topological transition in the bulk of a non-Hermitian system
%1.topological transition, waveguide array

\bibitem {Weimann2017}S. Weimann, M. Kremer, Y. Plotnik, Y. Lumer, S. Nolte,
K. G. Makris, M. Segev, M. C. Rechtsman, and A. Szameit, Nat. Mater.
\textbf{16}, 433 (2017).
%Shiyan:(Topologically protected bound states in photonic parity?Ctime-symmetric crystals)
%1.topological transition,photonic crystals


\bibitem {Bandres2018}M. A. Bandres, S. Wittek, G. Harari, M. Parto, J. Ren,
M. Segev, D. N. Christodoulides, and M. Khajavikhan, Science \textbf{359},
4005 (2018).
%Shiyan:(Topological insulator laser: Experiments)
%1.topological edge states, laser

\bibitem {Zhou20182}H. Zhou, C. Peng, Y. Yoon, C. W. Hsu, K. A. Nelson, L. Fu,
J. D. Joannopoulos, M. Soljacic, and B. Zhen,
Science \textbf{359}, 1009 (2018).
%Shiyan:(Observation of bulk Fermi arc and polarization half charge from paired exceptional points)
%EPs, photonic crystals

\bibitem {Cerjan2019}A. Cerjan, S. Huang, M. Wang, K. P. Chen, Y. Chong, and
M. C. Rechtsman, Nat. Photon. \textbf{13}, 623 (2019).
%Shiyan:(Experimental realization of a Weyl exceptional ring)
%EPs, photonic crystals
\bibitem {Xiao2017}L. Xiao, X. Zhan, Z. H. Bian, K. K. Wang, X. Zhang, X. P.
Wang, J. Li, K. Mochizuki, D. Kim, N. Kawakami, W. Yi, H. Obuse, B. C.
Sanders, and P. Xue, Nat. Phys. \textbf{13}, 1117 (2017).
%Shiyan:(Observation of topological edge states in parity?Ctime-symmetric quantum walks)
%1.topological edge states, quantum walks

\bibitem {Wang2019}K. Wang, X. Qiu, L. Xiao, X. Zhan, Z. Bian, B. C. Sanders,
W. Yi, and P. Xue, Nat. Commun. \textbf{10}, 2293 (2019).
%Shiyan:(Observation of emergent momentum?Ctime skyrmions in parity?Ctime-symmetric non-unitary quench dynamics)
%1.topological , quantum walks

\bibitem {Xiaoxue2019}L. Xiao, T. Deng, K. Wang, G. Zhu, Z. Wang, W. Yi, P.
Xue, Nat. Phys. \textbf{16}, 761 (2020).
%Shiyan:(Observation of non-Hermitian bulk-boundary correspondence in quantum dynamics)(7)
%1.BBC, quantum walks

\bibitem {Helbig2019}T. Helbig, T. Hofmann, S. Imhof, M. Abdelghany, T.
Kiessling, L. W. Molenkamp, C. H. Lee, A. Szameit, M. Greiter, and R. Thomale,
Nat. Phys. \textbf{16}, 747 (2020).
%Shiyan:(Observation of bulk boundary correspondence breakdown in topolectrical circuits)(7)
%1.BBC, circuits

\bibitem {Wang2015}X. Wang, T. Liu, Y. Xiong, and P. Tong, Phys. Rev. A \textbf{92},
012116 (2015).
%Chaodao: Spontaneous PT -symmetry breaking in non-HermitianKitaev and extended Kitaev models[J].

\bibitem {San2016}P. San-Jose, J. Cayao, E. Prada, and R. Aguado, Sci. Rep.
\textbf{6}, 21427 (2016).
%chaodao (Majorana bound states from exceptional points in non-topological superconductors)????


\bibitem {Yuce2016}C. Yuce, Phys. Rev. A \textbf{93}, 062130 (2016).
%Chaodao: Majorana edge modes with gain and loss


\bibitem {Zeng2016}Q. B. Zeng, B. Zhu, S. Chen, L. You, and Rong L$\ddot{u}$, Phys. Rev. A
\textbf{94}, 022119 (2016).
%Chaodao: Non-Hermitian Kitaev chain with complex on-site po-tentials[J]. Phys. Rev. A, \textbf{92},012116 (2015). Phys. Rev. A, \textbf{94}, 022119 (2016). Phys. Rev. B, \textbf{98}, 085116 (2018)


\bibitem {Menke2017}H. Menke, M. M. Hirschmann, Phys. Rev. B \textbf{95},
174506 (2017).
%Chaodao: Topological quantum wires with balanced gain and loss[J].

\bibitem {Kawabata2018}K. Kawabata, Y. Ashida, H. Katsura, and M. Ueda, Phys.
Rev. B \textbf{98}, 085116 (2018).
%chaodao(Parity-timesymmetric topological superconductor)

\bibitem {Lieu2019}S. Lieu, Phys. Rev. B \textbf{100}, 085110 (2019).
%Chaodao: Non-Hermitian Majorana modes protect degenerate steady states

\bibitem {Li2018}C. Li, X. Z. Zhang, G. Zhang, and Z. Song, Phys. Rev. B
\textbf{97}, 115436 (2018).
%Chaodao: Topological phases in a Kitaev chain with imbalanced pairing[J].


\bibitem{SongZhi} X. Z. Zhang and Z. Song, Ann. Phys. \textbf{339}, 109 (2013). %Momentum-independent reflectionless transmission in the non-Hermitian time-reversal symmetric system.

 \bibitem{ImblanceParing} P. Matthews, P. Ribeiro, and A. M. Garc\'{\i}a-Garc\'{\i}a, Phys. Rev. Lett. \textbf{112}, 247001 (2014)
     %Dissipation in a Simple Model of a Topological Josephson Junction.
\bibitem {Unpair2019}K. Yamamoto, M. Nakagawa, K. Adachi, K. Takasan, M. Ueda, and N. Kawakami, Phys. Rev. Lett. \textbf{123}, 123601 (2019)
\bibitem{HemitianKC} M. Franz, Nat. Nanotechnol. \textbf{8}, 149 (2016).
\bibitem{PTKC} Y. Ashida, S. Furukawa, and M. Ueda, Nat. Commun. \textbf{8}, 15791 (2017).
\bibitem{Array} A. McDonald, T. Pereg-Barnea, and A. A. Clerk Phys. Rev. X \textbf{8}, 041031 (2018)
\bibitem{PwaveParing} L. Jiang, T. Kitagawa, J. Alicea, A. R. Akhmerov, D. Pekker,
G. Refael, J. I. Cirac, E. Demler, M. D. Lukin, and P. Zoller,
Phys. Rev. Lett. \textbf{106}, 220402 (2011).
\bibitem{NHKC1} T. E. Lee and C. K. Chan, Phys. Rev. X \textbf{4}, 041001 (2014);
\bibitem{NHKC2}T. E. Lee, F. Reiter, and N. Moiseyev, Phys. Rev. Lett. \textbf{113}, 250401
(2014).
\bibitem{NHKC3} Y. Ashida, S. Furukawa, and M. Ueda, Phys. Rev. A \textbf{94}, 053615 (2016).
\bibitem{NHKC4} J. Li, A. K. Harter, J. Liu, L. de Melo, Y. N. Joglekar, and L. Luo, Nat. Commun. \textbf{10}, 855 (2019).
\bibitem{JWTran1} E. H. Lieb, T. Schulz, D. C. Mattis, Ann. Phys. \textbf{16}, 407 (1961)%Two soluble models of an antiferromagnetic chain.
\bibitem{JWTran2} D. C. Mattis. Phys. Today \textbf{39}, 62 (1986).
%: The Theory of Magnetism II
\bibitem {Guo2016}J. S. Xu, K. Sun, Y. J. Han, C. F. Li, J. K. Pachos, and G.
C. Guo, Nat. Commun. \textbf{7},13194 (2016)
\bibitem{SemiFinit} S. Hegde, V. Shivamoggi, S. Vishveshwara, and D. Sen, New Journal of Physics \textbf{17}, 053036 (2015)  %Parity blocking in quenching dynamics of Majorana wires.

\bibitem{SemiFinit2} S. S. Hegde, S. Vishveshwara, Phys. Rev. B, \textbf{94}, 115166 (2016). % . Majorana wave-function oscillations, fermion parity switches, and disorder in Kitaev chains[J].
\bibitem{SemiFinit3}N. Leumer, M. Marganska, B. Muralidharan, and M. Grifoni, J. Phys.: Condens. Matter \textbf{32},445502 (2020).
    %Analytical spectrum and the sparsity of Majorana zero modes in the topological phase diagram of the finite Kitaev chain[J]. 2019. Exact eigenvectors and eigenvalues of the finite Kitaev chain and its topological properties.


\end{thebibliography}
\end{document}